# DIC2Abaqus: Calculating mixed-mode stress intensity factors from 2D and 3D-stereo displacement fields

Abdalrhaman Koko [1a,b] and T. James Marrow [b]

[a] National Physical Laboratory, Hampton Road, Teddington TW11 0LW, United Kingdom

[b] Department of Materials, University of Oxford, Oxford OX1 3PH, United Kingdom

## Abstract

Evaluating the conditions for crack propagation under static and cyclic loads is critical for predicting the lifespan of engineering components, particularly in the energy and transport industries. Digital Image Correlation (DIC) provides precise displacement field measurements that can be used to calculate strain energy release rates and stress intensity factors (SIFs), but integrating DIC data into computer-aided engineering (CAE) software like Abaqus, a widely used finite element package, remains challenging. This paper introduces *DIC2Abaqus*, a freely available MATLAB-based tool that automates DIC data processing in Abaqus to extract material properties in isotropic and anisotropic elastic and elastoplastic materials. It employs the *J*-integral and interaction integral methods to compute mixed-mode SIFs, including mode III, without requiring a predefined specimen geometry or applied loads. It supports 2D and 3D-stereo DIC data and streamlines the process from geometry creation to job submission and post-processing. Validation against analytical and experimental results confirms its accuracy and reliability. By taking fracture mechanics analyses beyond ISO and ASTM standards, *DIC2Abaqus* offers a versatile, efficient, and accessible simulation tool for industry, research, and education. The source code and tutorials can be freely downloaded from https://shorturl.at/p47rI.

---

[1] Corresponding author. E-mail address: abdo.koko@npl.co.uk

# 1. Introduction

Fracture mechanics focuses on understanding and predicting the propagation of cracks, which involves quantifying the conditions required for crack growth [1–5]. A key concept is the strain energy release rate, representing the potential energy to extend a crack by creating new surfaces. In linear elastic materials or under small-scale yielding (SSY) conditions, the strain energy release rate can be linked to the stress intensity factor (SIF), a descriptor of the magnitude of the stress field around a crack [6–8]. Even when crack tip plasticity invalidates SSY, the strain energy release rate remains a valuable metric [9–11] that describes the conditions at the crack tip. These principles, therefore, also govern other important fracture mechanisms, including fatigue and stress corrosion cracking [12,13].

Traditionally, SIFs and strain energy release rates have been computed using analytical solutions [14,15] or finite element methods (FEM) based on applied loads, boundary conditions, and specimen geometry [16,17]. However, real-world scenarios introduce uncertainties that can undermine these standard approaches. Residual stresses from manufacturing [18,19], frictional effects [20], specimen misalignment [21,22], or unknown boundary conditions complicate the accurate characterisation of the crack stress field. Additionally, analytical solutions assume idealised conditions that may not hold in practical cases. For example, when a crack deviates from its expected path due to kinking or branching or changes in load alignment, the assumptions underlying the analytical calculations break down, leading to errors in the experimentally derived fracture parameters.

This limitation is particularly relevant in complex engineering components, which may have poorly defined loading conditions and complex crack paths. To address these issues, there is growing interest in directly measuring deformation fields around cracks [23–29]. These direct measurements can provide a more correct assessment of the critical conditions for fracture.

Evaluation of the SIF is necessary to quantify material properties. As noted above, these are calculated conventionally using known boundary conditions (loads or displacements) and standard specimen geometries. However, challenges arise, such as fatigue crack closure, uncertain boundary conditions (e.g., due to experimental inaccuracies), or complex specimen geometries that can interfere with these calculations. Additionally, uncertainty from human error, assuming minimal test system compliance, and load misalignment before or during the test can influence the results significantly, which makes the use of analytical solutions incorrect [30,31]. Significantly, comparisons between the International Organization for Standardization (ISO) and the American Society for Testing and Materials (ASTM) standards for measuring fracture toughness have reported between 58% [32] – 24% [33] experimental disparity and 5% theoretical inconsistency [34] between the two standards. Consequently, methods for direct evaluation of the crack field, such as digital image correlation (DIC), have gained interest [35–39].

DIC has become a widely adopted technique for full-field displacement measurement due to its versatility and ease of application across different materials [36,40–42]. DIC provides precise displacement maps by tracking surface patterns between sequential images, making it highly valuable for fracture mechanics studies [40,43–45]. Both 2D and 3D-stereo DIC (Figure 1) setups are used to capture in-plane and out-of-plane displacements, respectively. However, direct extraction of fracture parameters from DIC data requires advanced post-processing techniques. Field fitting approaches, such as least-squares optimisation using Williams' series [46], have been applied but can be quite sensitive to crack tip localisation [47–49]. An alternative finite element analysis (FEA)-based $J$-integral method directly computes the strain energy release rate and SIFs from the measured displacement field and is robust to uncertainties in crack tip positioning [24,48,50,51].

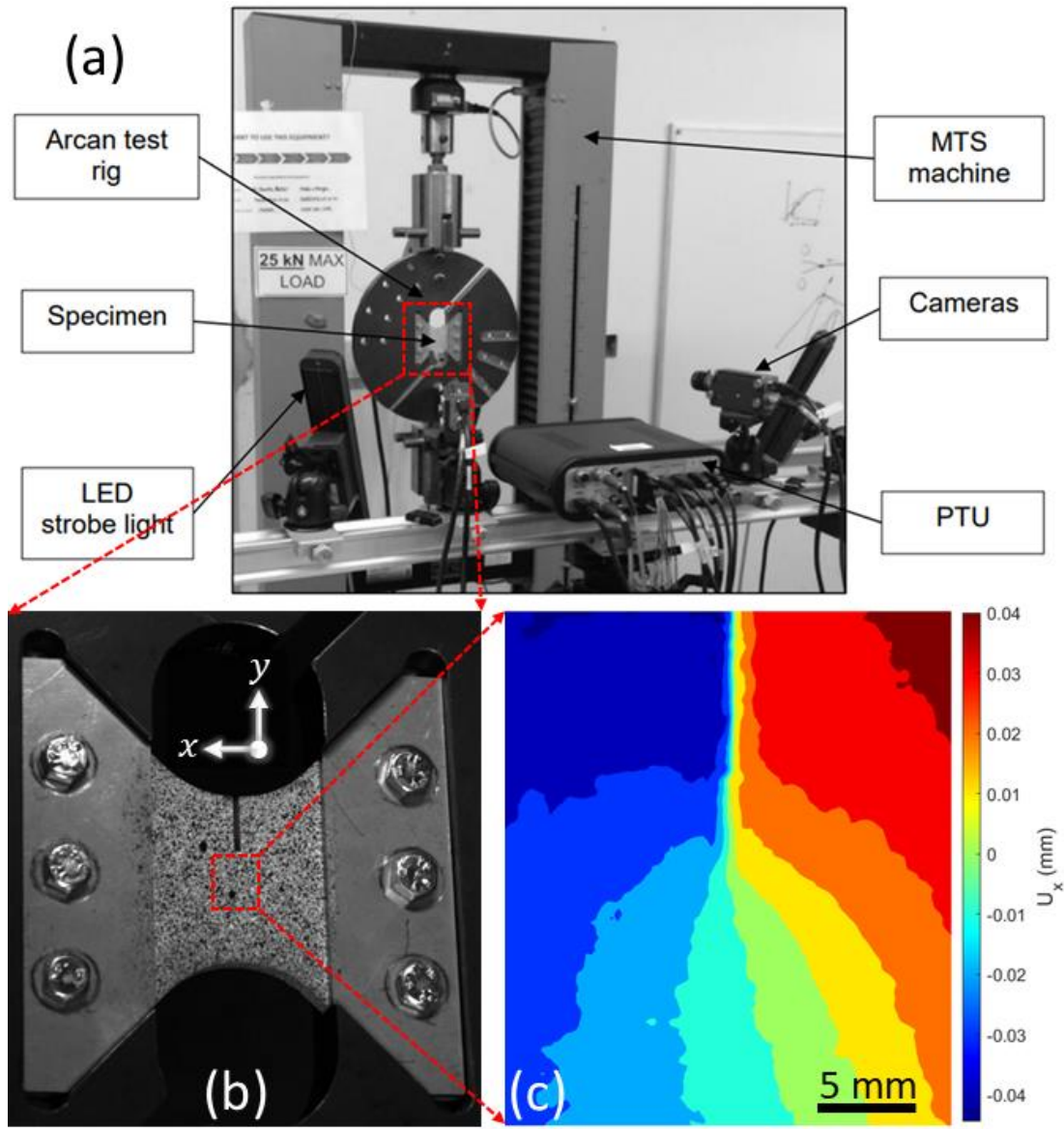

Figure 1: (a) An Arcan test fixture with a speckled butterfly sample and stereo digital imaging system. (b) Image of the field of view of the loaded speckled butterfly sample. (c) The $U_x$ displacement field is obtained by digital image correlation of the speckled field of view.

FEA provides a robust framework for processing experimental displacement fields. Standard FEA software, such as Abaqus, incorporates domain-integral methods to compute *J*-integrals and stress intensity factors, which is advantageous for SIF calculation compared to conventional analytical techniques [52,53]. However, integrating experimentally measured displacement data into FEA remains challenging due to differences in mesh structures and interpolation requirements. Some existing solutions rely on modifications of DIC algorithms [37,54], such as finite element-based enrichment techniques [55,56], which are not widely implemented in practice.

Recent advancements in full-field measurement methods (Material Testing 2.0 [57]), combined with increased computational power, high-fidelity multiscale modelling tools, and the application of machine learning and artificial intelligence (AI), have unlocked substantial potential across materials science and engineering applications. By leveraging full-field measurements, it will be possible to develop multiscale and data-driven models that accurately understand the process-structure-property (PSP) relationship, enabling precise evaluation of material limitations and optimised materials design.

We introduce *DIC2Abaqus*, a freely available MATLAB-based tool that automates the conversion of DIC displacement data into an Abaqus-compatible format. This enables the direct integration of experimental DIC measurements into FEA, allowing the extraction of fracture parameters for isotropic and anisotropic elastic and elastoplastic materials. Using the J-integral and interaction integral methods, *DIC2Abaqus* computes mixed-mode SIFs, including mode III, without requiring predefined specimen geometry or applied loads. It accommodates complex geometries and non-standard crack shapes, making it highly adaptable to real-world applications. *DIC2Abaqus* streamlines the workflow from experimental measurement to numerical simulation. Its accessibility and ease of use make it a valuable tool for industry, research, and education, enhancing the predictive capabilities of computational models in materials science and engineering. Moreover, although the focus of this paper (and its chosen examples) is on measured displacement fields, the DIC2Abaqus code can also be applied to displacement fields that have been calculated by integration of measured elastic strain fields (i.e. from diffraction) to estimate the equivalent elastic displacement fields [58,59].

## 2. Structure of the software

The software, implemented in MATLAB, interacts with Abaqus to calculate the *J*-integral and SIFs directly from the displacement field measured using DIC. This includes stereo-DIC (Figure 1a and b), where two cameras are used to measure not only the in-plane displacement fields but also the out-of-plane displacement field. As shown in Figure 2, the software is organised into several key modules that handle user input, data pre-processing, *J*-integral and SIF calculations, and postprocessing. This modular framework enhances flexibility, allowing users to adapt the software to various complex scenarios, including the detailed study of crack propagation and material behaviour under stress.

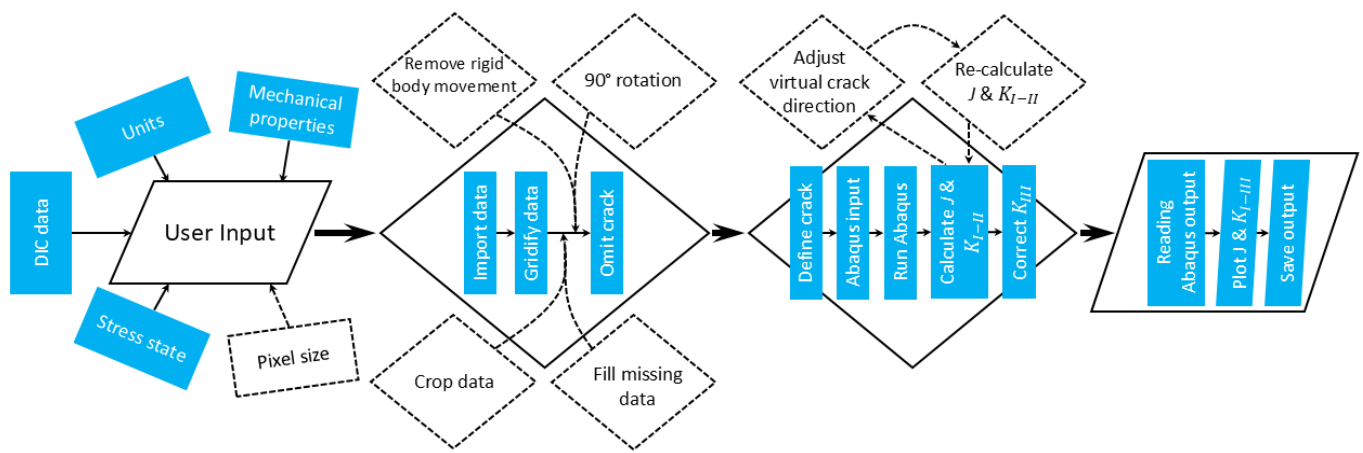

Figure 2: *DIC2Abaqus* structure. Dashed lines indicate optional functions.

### 2.1. User input

The software requires the user to input the location of the DIC data, the DIC units (meters, millimetres, or micrometres), and the material's mechanical properties, which can be an isotropic or anisotropic[2] elastic material, or an elastoplastic material that follows the Ramberg–Osgood relationship [60] as defined below:

$$E\varepsilon = \sigma + \alpha\,\sigma\left(\frac{\sigma}{\sigma_0}\right)^{n-1} \qquad (1)$$

---

[2] The user can input the material anisotropic modulus which can be used to calculate the effective Young's modulus, shear modulus and Poisson's ratio for cubic anisotropic materials based on ref. [100]. For non-cubic crystals, the user needs to input the elastic constants. This information will be needed later to calculate the SIFs from the strain energy release rate.

where $\varepsilon$ and $\sigma$ are the stress and strain, $\sigma_0$ is the yield stress, $\alpha$ is the yield offset, and $n$ hardening exponent (>1).

The input data can be uploaded in any format that MATLAB can read. The data needs to have four columns for 2D-DIC data: X-coordinate, Y-coordinate, displacement in X and displacement in Y for each measurement point. For stereo-DIC data (six columns), the first three columns are for each measurement point X, Y and Z-coordinates and the others record the displacements in X, Y, and Z, i.e. $U_x$, $U_y$, and $U_z$, respectively. Currently, the experimental DIC data need to be in a regularised grid to be transformed into a format that can be utilised for simulations and analyses in Abaqus.

The user is asked to locate and identify the crack in the displacement field to be analysed, with options of rotating and cropping the field of view. After preprocessing, the function moves on to the crack detection step. For straight cracks, the user selects the crack tip and crack mouth and then chooses the region around the crack that needs to be masked interpolation, as shown in Figure 3. The purpose of this mask is to exclude poor quality or uncertain data, which DIC commonly provides close to discontinuities [61–63]. The displacements within this region are later obtained by finite-element-based interpolation.

For tortuous cracks, the crack geometry – including poor quality or uncertain DIC data – must be excluded beforehand from the DIC data. This can be done manually or autonomously by incorporating an algorithm, such as [39,61,62,64], to detect and exclude the crack. Next, the user can input the crack tip coordinates and direction/angle or select these when promoted. This information will be used to build the finite element model in Abaqus, including the crack direction (q-vector illustrated in Figure 4), which is assumed to be parallel to the defined crack.

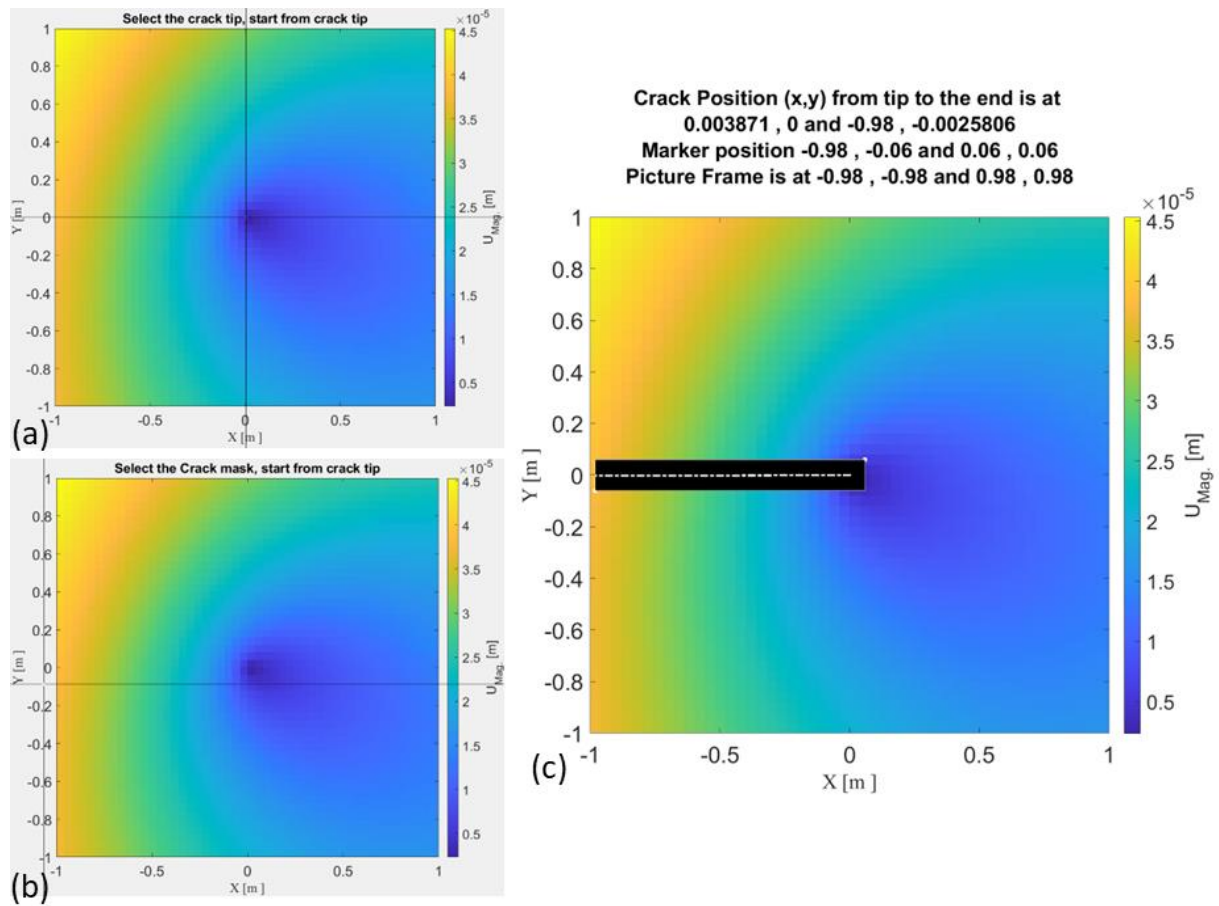

Figure 3: After loading the displacement field, the script visualises the displacement magnitude (U$_{Mag}$) to facilitate user interaction. (a) The user selects the crack path starting from the crack tip. (b) The user defines a masking region to exclude areas with poor data. (c) The selected crack path (dotted white lines) and (d) the masked region (black square) are displayed on the U$_{Mag}$ map.

## 2.2. Pre-processing

One of the key steps is the generation of Python code that Abaqus executes automatically. This step converts processed DIC data into a format compatible with the Abaqus environment. It creates input files (.inp for tortuous crack) or scripts that define the simulation setup, including material properties, crack geometry, meshing, and boundary conditions.

The process also includes a step for managing simulation output, which defines how results are stored and accessed for analysis and interpretation. This generates output files and structured data storage for efficient result retrieval. Logging and documentation mechanisms are implemented to track the simulation process, recording details such as errors or warnings during code generation and execution to provide a reference for debugging and future use.[3]

### 2.3. *J*-integral and SIF calculations

The software deploys Abaqus in the background to calculate the *J*-integral from the displacement data provided by DIC, as shown in Figure 1c. A Python script, which is automatically written using the software, handles the creation of a finite element model in Abaqus that aligns with the DIC data, ensuring both share the same coordinate system and nodes. Matching the finite element mesh with the DIC grid avoids interpolation.

The script uses a rectangular grid with four nodes per plane stress/strain element (Abaqus' CPS4/CPE4). The DIC-measured node displacements are imposed as boundary conditions, except in the masked region near the crack. The Abaqus software then applies the chosen material laws, and the *J*-integral is calculated using the equivalent domain integral (EDI) method with a Virtual Crack Extension (VCE) technique [53,65,66]. This method calculates the potential elastic strain energy release rate. The EDI method starts at the crack tip and propagates in the direction of a VCE, with a smooth function (q) varying from unity at the tip to zero at the outer domain. The q-vector (Figure 4) is normal to the crack front, aligning with the crack path. Multiple contours are used in the *J*-integral calculation to ensure contour independence, determining the potential release of elastic strain energy from crack propagation based on the displacement field (Figure 4).

The singular integral can be directly related to the SIFs through mode-decomposition techniques, such as the interaction integral method [67,68]. The interaction integral method – implemented in Abaqus – is particularly useful for evaluating mixed-mode fracture problems, as it enables the calculation of SIFs for each mode independently, even in complex, non-uniform stress fields. Unlike conventional approaches that rely on direct field fitting, the interaction integral leverages auxiliary fields to isolate individual mode contributions while maintaining accuracy in cases of non-proportional loading or irregular crack geometries, providing a more comprehensive understanding of crack-tip mechanics [69–71]. In linear elastic fracture mechanics (LEFM), the elastic strain field is typically decomposed into three primary modes: mode I (tensile/compression), mode II (in-plane shear), and mode III (out-of-plane shear). The interaction between these modes is critical in determining the crack propagation direction [72–74].

---

[3] Note that for Linux users, a modification is required in the code at line 42, with an alternative line already commented out at line 43 in the *PrintRunCode* function. Additionally, the directory separator must be adjusted from "/" to "\" within the file naming process to ensure compatibility with Linux systems

For the stereo-DIC field with anisotropic or isotropic elastic material containing the out-of-plane displacement field ($U_z$), the mode III SIF is calculated by injecting the field as $U_x$ displacement. The antisymmetric (pseudo) shear stress intensity factor ($K_{II}^{\text{pseudo}}$) is calculated by Abaqus, and the pseudo-in-plane shear mode II SIF is then corrected to an out-of-plane mode III SIF as described in equation (2), where $E$ is Young modulus and $G$ is the shear modulus. The total J-integral ($J$) is then calculated by summing the mode I and II J-integral ($J^{I,II}$) and the mode III J-integral ($J^{III}$), as shown in equation (3).

$$K_{III} = \frac{2G}{E} K_{II}^{\text{pseudo}} \qquad (2)$$

$$J = J^{I,II} + J^{III} \qquad (3)$$

It should be highlighted that the sign (positive and negative) of the in-plane shear (II) and out-of-plane shear (III) components is not significant in this scenario, as it is determined by the nodal configuration at the crack tip and lacks physical relevance. In contrast, the sign of mode I is critical, as it indicates whether tensile (positive) or compressive (negative) stresses are acting at the crack tip. The symmetric out-of-plane component of mode I remains unaffected by the orientation of the mode III stress intensity factor.

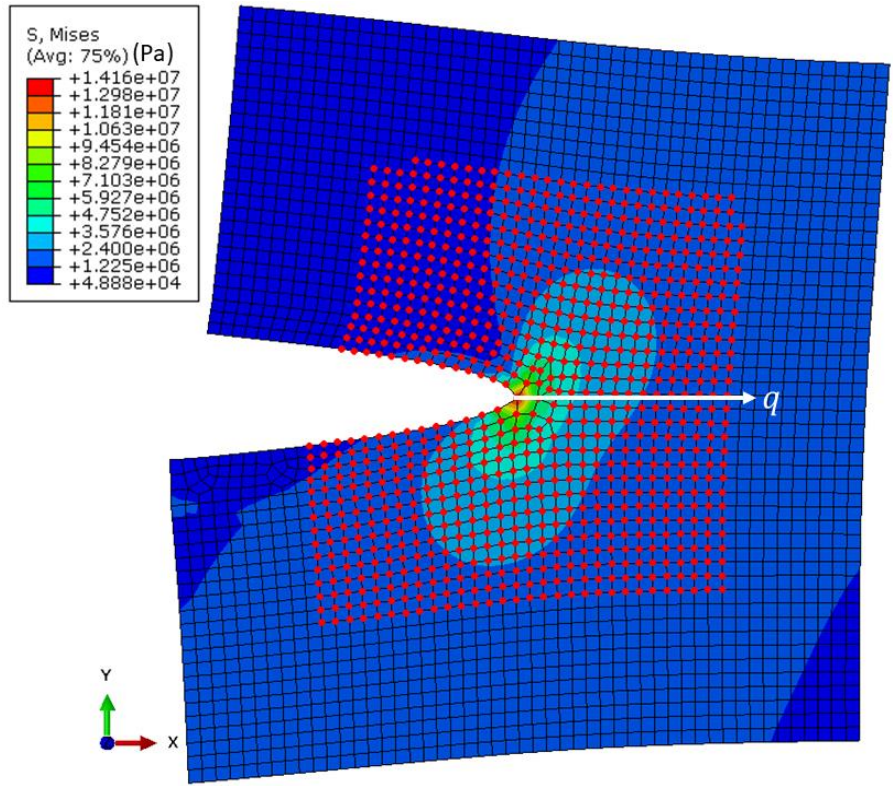

Figure 4: Deformed configuration of the von Mises stress around the crack calculated from the DIC displacement field. The red nodes donate the domain to be integrated, or equivalent domain integral (EDI), for the *J*-integral calculation, and the q-vector is VCE. The domain starts from the crack tip and expands incrementally to engulf the entire field of view, and the *J*-integral and SIFs are calculated as a function of EDI.

## 2.4. Post-processing

After completing the Abaqus simulation, the software extracts and prepares output data, including the *J*-integral and mode I-III SIFs, for visualisation. After reading the Abaqus output files (ODB and text-based formats), the data is parsed to identify and structure relevant parameters such as the *J*-integral and SIFs. The user is prompted to select the number of contours to consider. Contributions from surrounding gradient fields can disrupt path independence unless fully enclosed within the integration domain [75–77], so convergence may fail when the integration domain extends into peripheral stress fields. The video in the supplementary material visualises convergence with the integration domain expansion.

The *J*-integral and SIF values over the expanding contours are processed, and the integration convergence region (shaded pink in Figure 6) is identified to compute the mean values and variance. The visualised results provide insights into crack behaviour, as illustrated in Figure 6.

Page 11 of 32

The workflow includes an optional adjustment step to refine the q-vector direction for *J*-integral calculations based on Abaqus output. This step helps determine the potential crack propagation direction and allows for iterative refinements based on user inputs. The computational process can be repeated with an updated crack direction to improve accuracy. For more details about the effect of the crack direction, see Appendix A.

## 3. Illustrative examples

Here, we illustrate the software's capabilities: i) using a synthetic data set to validate the code, and ii) using a scanning electron microscope (SEM) DIC field of a crack with complicated geometry to highlight the versatility of the software. More examples are provided in the *InputDesk_Validate function,* and an example of how to use the code with typical DIC data is provided in the *InputDesk_DIC* function of the code.

### 3.1. Straight crack and validation of the software

A synthetic displacement field for a mixed-mode crack in an infinite body was created using an analytical solution based on the desired stress intensity factor [78]. The field has a mode I stress intensity factor ($K_I$) of 3 MPa m$^{0.5}$, mode II ($K_{II}$) of 1 MPa m$^{0.5}$, and mode III ($K_{III}$) of 5 MPa m$^{0.5}$, and plane strain conditions (equation 4). The elastic modulus ($E$) and Poison's ratio ($v$) were 210 GPa and 0.3, respectively. The data are presented in the field of view of 50 x 50 elements ( 2 x 2 µm²), with 0.04 x 0.04 µm² square elements, and the crack tip was placed at the centre (0,0).

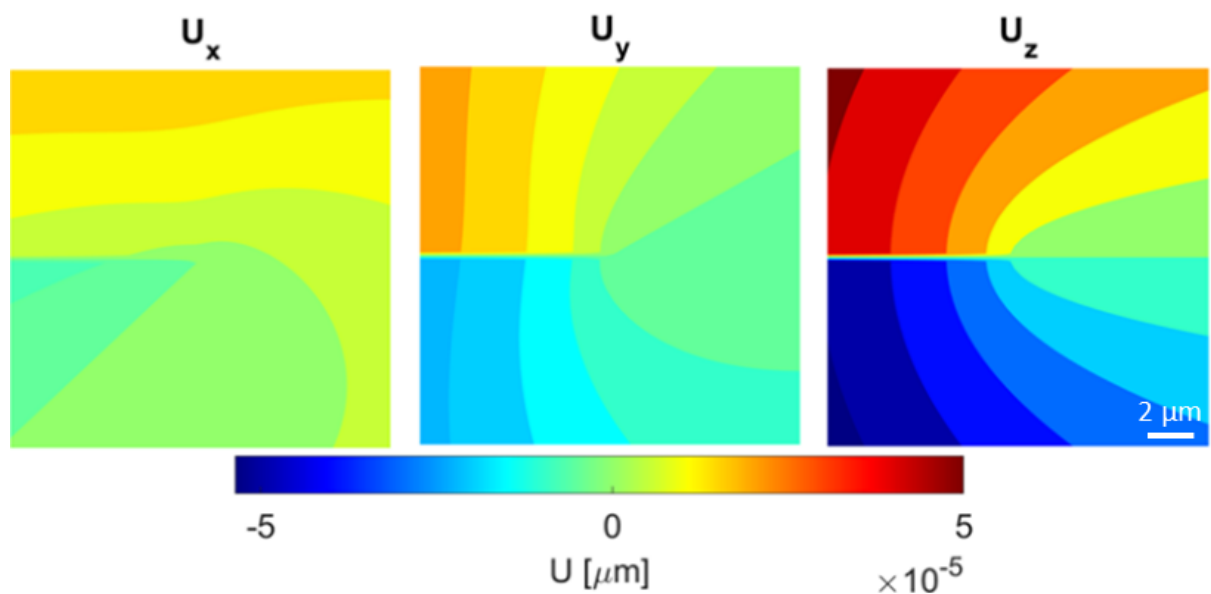

Figure 5: Synthetic $U_x$, $U_y$, and $U_z$ displacement field components with the crack tip at the centre.

$$\begin{Bmatrix} U_x \\ U_y \\ U_z \end{Bmatrix} = \frac{1}{\mu}\sqrt{\frac{r}{2\pi}} \begin{bmatrix} \cos\frac{\theta}{2}\left(1 - 2v + \sin^2\frac{\theta}{2}\right) & \sin\frac{\theta}{2}\left(2 - 2v + \cos^2\frac{\theta}{2}\right) & 0 \\ \sin\frac{\theta}{2}\left(2 - 2v - \cos^2\frac{\theta}{2}\right) & \cos\frac{\theta}{2}\left(-1 + 2v + \sin^2\frac{\theta}{2}\right) & 0 \\ 0 & 0 & 2\sin\frac{\theta}{2} \end{bmatrix} \begin{Bmatrix} K_I \\ K_{II} \\ K_{III} \end{Bmatrix} \quad (4)$$

$$\text{Shear modulus } (\mu) = \frac{E'}{2(1 + v)}, \quad E' = \frac{E}{1 - v^2}$$

Synthetic $U_x$, $U_y$, and $U_z$ displacement fields around the stationary mixed-mode crack (Figure 5) were used to calculate mode I SIF, in-plane asymmetrical mode II SIF, and out-of-plane asymmetrical mode III SIF (Figure 6). The EDI method implemented in Abaqus was used on the decomposed fields, and stabilised convergence was achieved as the domain expanded, with the calculated *J*-integral and decomposed stress intensity factors matching the values used as inputs to create the field, with the deviation being due to the manual selection of the crack tip. For more details on the error analysis, please refer to Appendix B.

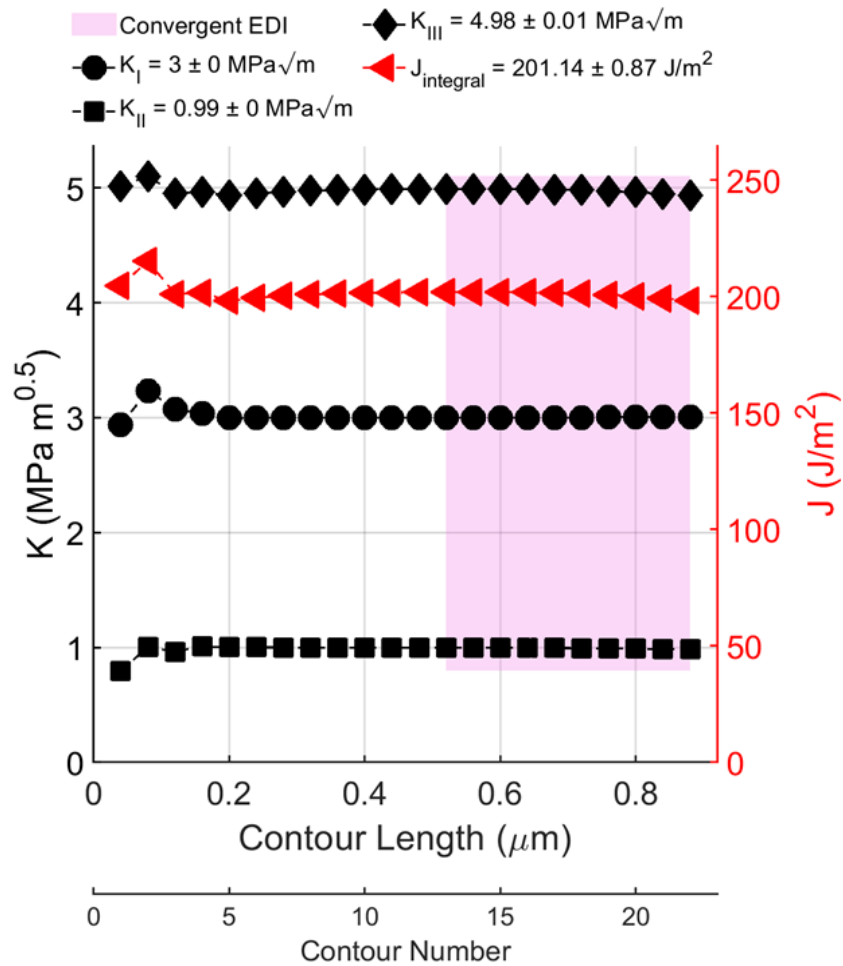

Figure 6: *J*-integral and decomposed loading modes as a function of contour distance from the crack tip for synthetic data of a crack experience 3 MPa.m$^{0.5}$ of mode I loading, 1 MPa.m$^{0.5}$ of mode II loading and 5 MPa.m$^{0.5}$ of mode III loading. The represented values are calculated from the highlighted pink area.

## 3.2. Tortuous cracks

Short cracks are microstructure-sensitive, typically mixed-mode, involve localised plasticity, and often have a complicated geometry [79–81]. Existing analysis of the microscopic stresses acting on short fatigue cracks lacks reliable quantitative measurement techniques due to the difficulty of conducting in situ experiments and the complexity of analysing or modelling short cracks [82]. These local analyses are valuable when the external conditions are unknown or uncertain, and, especially at the micro-scale, they provide alternatives to the existing methods and analytical solutions that use micro-pillars, micro-cantilevers, and indentation [83–85].

Here, we look at the in-plane displacement field measured in an in-situ experiment conducted inside a scanning electron microscope (SEM) for a compact tension sample made of aluminium 5052 alloy. The sample was loaded in tension and was speckled using gold nanoparticles to enable DIC (Figure 7a). More details about the experiment can be found in [86].

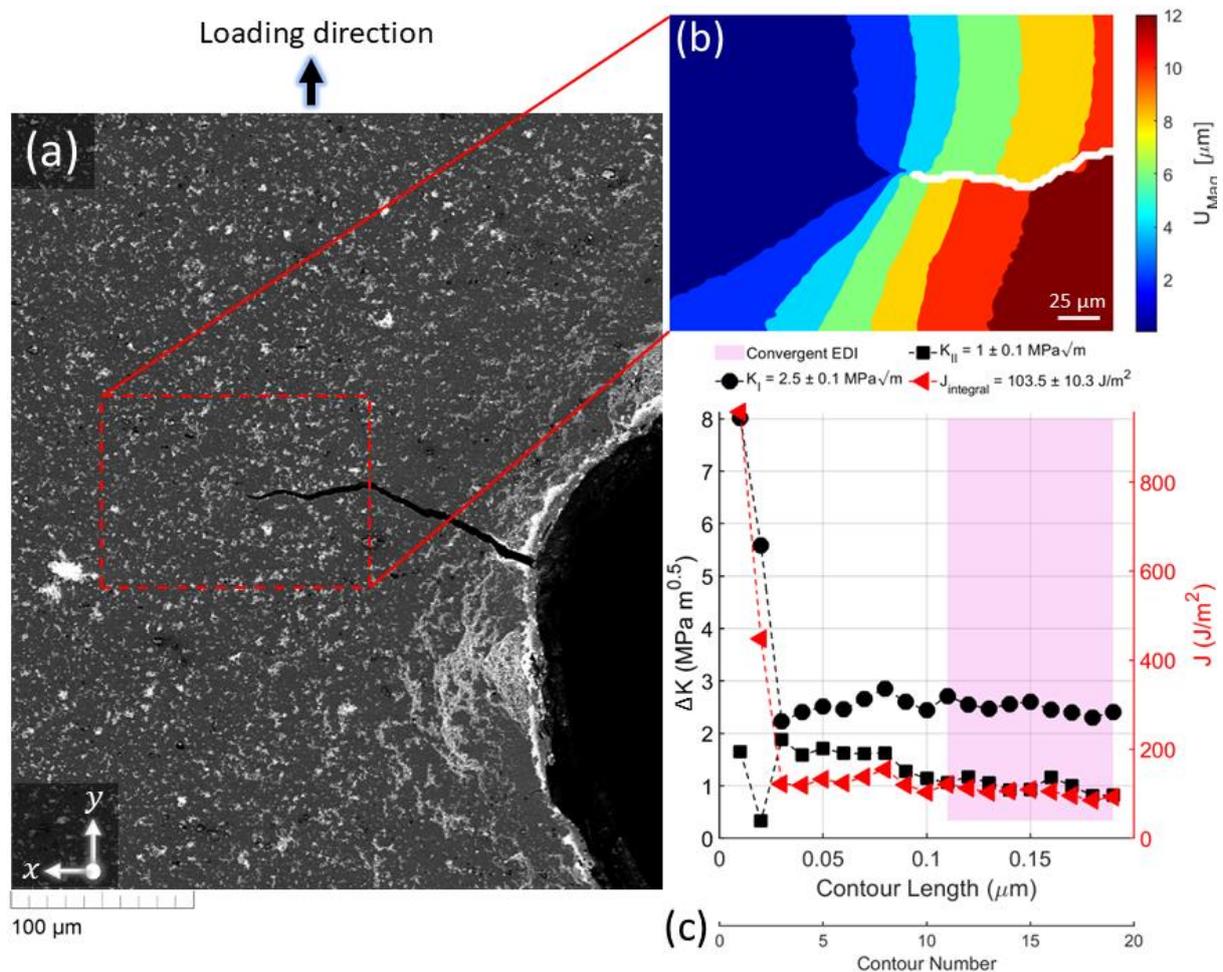

Figure 7: (a) Curved crack in an aluminium 5052 alloy compact tension sample. (b) The magnitude of the experimental displacement field around the crack (white). (c) Calculated *J*-integral and mode I-II SIFs from the displacement field.

The crack geometry was determined using the phase congruency of the displacement field (Figure 7b) [64,87], and excluded from the in-plane displacements. The SEM-DIC field was then used in Abaqus via *DIC2Abaqus*, and the effective[4] elastic strain energy release rate was calculated along with the mode I and II SIFs (Figure 7c). The initial contours were non-convergent due to the highly localised fields close to the crack tip [25,88,89], and stable convergence was achieved ~15 μm away from the crack tip, as path-integrals need to engulf stress concentrators to adequately describe them [90].

---

[4] The word "effective" is used because the displacement field was calculated between two consecutive images that were both captured with the crack being loaded.

Compared to the analytical solution based on ASTM E1820 [14], which assumes only mode I conditions exist at the crack tip, the measured field gives an entirely different picture; the crack is experiencing mixed mode conditions with mode I ($\Delta K_I$) of 2.5 ± 0.1 MPa.m$^{0.5}$ and mode II ($\Delta K_{II}$) of 1.0 ± 0.1 MPa.m$^{0.5}$, with elastic *J*-integral ($\Delta J_e$) of 103.5 ± 10.3 J/m$^2$. If the effect of crack tip plasticity is considered by applying an elastoplastic Ramberg–Osgood relationship with a yield stress of 193 MPa, 0.60 yield offset, and 8.87 hardening exponent estimated from the tensile testing, the strain energy release rate is 98.7 ± 17.4 J/m$^2$, which indicate minimal or no plasticity at the crack tip.

If transformed to align with the grain's orientation as determined through electron backscatter diffraction [91,92], the anisotropic stiffness matrix can be utilised to achieve more accurate calculations of mode I and II [91,92].

## 4. Conclusions

Material Testing 2.0 (MT2.0) is an opportunity to leverage full-field measurement methods and advanced computational tools to improve materials testing and design. The *DIC2Abaqus* tool provides a robust and efficient solution for integrating Digital Image Correlation (DIC) data into the FEA software Abaqus. The tool enables direct incorporation of experimental data into finite element models, which is critical where the sample does not fit standard methods. By automating the process of converting DIC data into Abaqus-compatible formats, the software enhances the accuracy of stress intensity factor (SIF) calculations even under complex conditions such as mixed-mode fractures and short cracks.

Future work aims to develop a similar tool for 3D digital volume correlation (DVC) [93].

## Appendix A: VCE direction effect

The q-vector is the virtual crack extension (VCE) direction [65] or the assumed direction that the crack will propagate to, which affects the calculation of the strain energy release rate or *J*-integral and mode I-III (Figure 8). Ideally, the q-vector should be in the direction of maximum strain energy release rate [86] or any other criteria typically used to predict crack direction [94–97]. In the *DIC2Abaqus*, for the straight crack, the crack direction is assumed to be parallel to the crack, and for the curved crack, the user selects the crack direction manually. Once the strain energy release is calculated in this direction, the user will be advised to change q to the direction with the maximum strain energy release rate ($q_{max}$) calculated [98,99]. To modify the $q_{max}$ criteria, the user can run the Python code that *DIC2Abaqus* writes and then manually, in the Abaqus interface, edit the crack propagation method to the maximum tangential stress criterion or $K_{II}=0$ criterion. See ABAQUS documentation for more details.

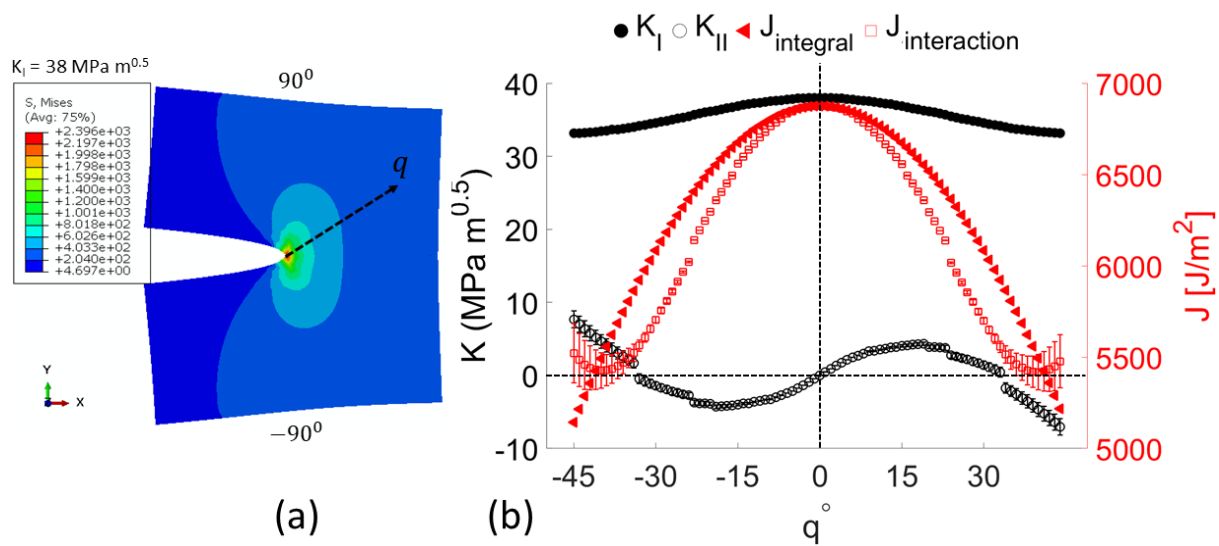

Figure 8: (a) ABAQUS® simulation of a two-dimensional crack on an isotropic material ($E$ = 210 GPa), loaded to 38 MPa.m$^{0.5}$ in mode I. (b) Calculated mode I and II while changing the assumed crack extension direction ($q$) using the interaction integral method (natively implemented in ABAQUS®). As the crack, it experiences increased mode II and reduced mode I.

# Appendix B: Uncertainty due to noise and crack tip position

The key sources of errors were assessed. First, a noise signal was randomly distributed to each displacement component. The magnitude of the signal increased from 0.0001% to 1% of the field mean magnitude, and the effect on the *J*-integral and SIFs was then studied. As shown in Figure 9, the EDI convergence decreased with increasing noise; hence, the uncertainty error in the value increased. The mean $K_I$ and *J*-integral values decreased, and the $K_{II}$ values increased, although $K_{III}$ was not adversely affected by the noise.

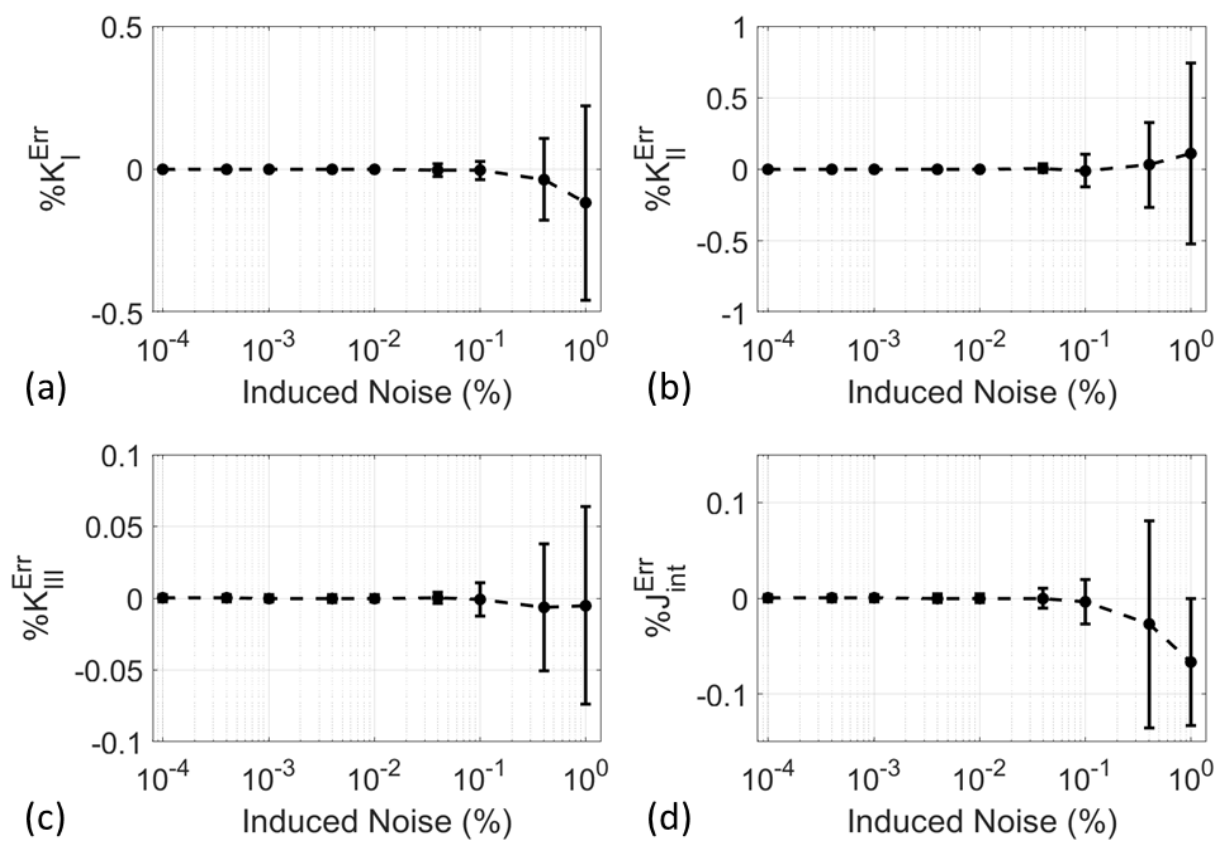

Figure 9: (a) $K_I$, (b) $K_{II}$, (c) $K_{III}$, and (d) *J*-integral normalised error analysis with induced random noise. The normalised error is calculated by taking the difference between the estimated value and the true value and dividing it by the true value.

Second, the sensitivity of the analysis to crack tip location was assessed relative to the accurate crack tip position, which was at the origin (0,0) coordinate. For *J*-integral calculation, the (normalised) error in locating the crack increases when the location is assumed to be ahead, compared to behind the actual crack tip location (Figure 10a). In $K_I$ and $K_{II}$ analysis (Figure 10b and c), the values are less affected by inaccuracy along the X-axis. However, for mode I, positional inaccuracies in the Y-axis increase the value when the crack was inaccurately located under its correct position and vice versa compared to mode II, which varies symmetrically as the tip deviates along the Y-axis. In $K_{III}$ analysis (Figure 10d), deviation from the tip is more complex but generally increases the value of $K_{III}$. The induced errors in each decomposed field combine and form the complex error seen in Figure 10a.

The overall trend of the SIFs is that uncertainties in the Y-axis position relative to the true position of the crack have more adverse effects than the X-axis. The sensitivity to the crack position does not affect the EDI convergence for SIF calculation (Figure 10e) but mainly affects the value. However, it affects both the convergence and the value for the *J*-integral.

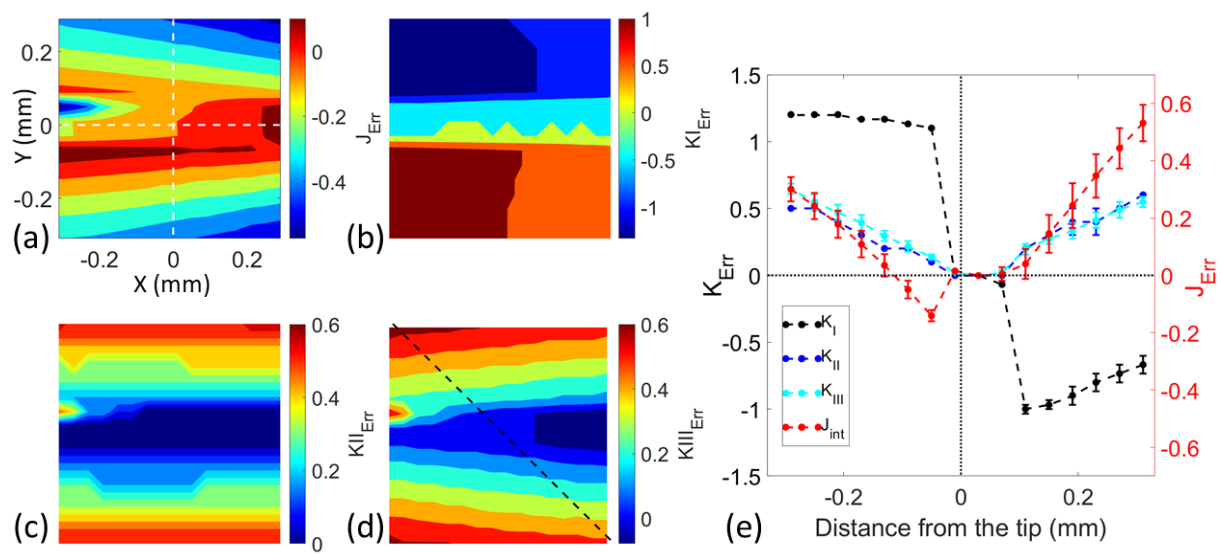

Figure 10: (a) J, (b) $K_I$, (c) $K_{II}$ and (d) $K_{III}$ normalised error analysis of the sensitivity to crack tip position originally at (0,0) highlighted by dashed white lines in (a). (e) The diagonal line profile (highlighted by the dashed black line in (d) of the *J*-integral and decomposed SIFs.

Overall, the proposed method was validated using the synthetic displacement field of a crack under mixed-mode loading, including the effect of noise and uncertainty of crack tip on the *J*-integral and SIFs values, which affect the EDI convergence and the *J*-integral and SIFs values, especially on the Y-axis (i.e., perpendicular to the crack direction).

## Acknowledgements

The authors thank Dr Louise Crocker and Dr Dalia Y. Ali for proofreading the article and the National Measurement System (NMS) programme of the UK government's Department for Science, Innovation and Technology (DSIT) for financial support.

## Code availability

A permanently archived version of the code can be found at https://doi.org/10.5281/zenodo.6411605, and a continuously updated version can be found at https://github.com/Shi2oon/DIC2ABAQUS.

## CRediT author statement

**Abdalrhaman Koko:** Methodology, Software, Visualisation, Investigation, Formal analysis, Writing - original draft.

**James Marrow:** Conceptualisation, Resources, Writing - review & editing.